\documentstyle[sprocl]{article}
\input{psfig}
\def\jou#1#2#3#4{{#1} {\bf #2}, #3 (#4)}
\def\annrev{\em Ann. Rev. Nucl. Part. Sci.}
\def\beq{\begin{equation}}
\def\eeq{\end{equation}}
\def\beqa{\begin{eqnarray}}
\def\eeqa{\end{eqnarray}}
\def\gsim{\mathrel{\rlap{\lower4pt\hbox{\hskip1pt$\sim$}}
    \raise1pt\hbox{$>$}}}         
\def\lsim{\mathrel{\rlap{\lower4pt\hbox{\hskip1pt$\sim$}}
    \raise1pt\hbox{$<$}}}         
\begin{document}
\title{HIGH-ENERGY GAMMA AND NEUTRINO ASTRONOMY\footnote{Invited talk 
at the 18th Texas      Symposium on Relativistic Astrophysics, to 
appear in the Proceedings  (eds A. Olinto, J. Frieman and D. Schramm, World
     Scientific, 1997) }}
\author{L. BERGSTR\"OM}
\address{Department of Physics, Stockholm University\\ Box 6730, S-113 85
Stockholm, Sweden }
\maketitle\abstracts{An overview is given of high-energy gamma-ray 
and neutrino astronomy, emphasizing the links between the two fields. 
With several new large detectors just becoming operational, the TeV 
gamma-ray and neutrino sky will soon be surveyed with unprecedented 
sensitivity.
 }
\long\def\comment#1{}
\def\cE{{\cal E}}

\section{Introduction }

These are exciting times for high-energy gamma ray and neutrino astronomy.
During the last couple of years several sources of TeV gamma rays have
finally been convincingly detected, after many years of marginal and 
sometimes erroneous claims of detection at higher energies in air shower 
arrays. 

This
healthy development of the field is due to the operation of several new
large experimental facilities, in particular the CASA, HEGRA and Whipple
experiments (for a summary of these experiments, see Ref.~\cite{croninetal}).  

In neutrino astronomy, the first sources beyond the Sun (and the transient
SN 1987A) remain to be discovered. There are great expectations that this
will happen soon, as new large neutrino telescopes are just about to
become operational.

There are several areas of intersection between gamma ray and neutrino
astronomy. By both probes one gets a view of violent astrophysical
processes, and in contrast to charged cosmic rays the direction to the source
is preserved. Most of the processes that give rise to high-energy neutrinos
should also generate gamma rays, and vice versa. By studying both types
of emission valuable information about the production mechanisms of these
energetic particles can be obtained. Due to the difference in 
absorption (TeV gamma rays are absorbed on IR intergalactic photons, 
whereas neutrinos are unaffected), useful information on the 
intergalactic radiation field may be obtained if far-away sources are 
observed.

Besides the more ``mundane'' local processes creating gamma rays and neutrinos,
such as cosmic ray collisions with interstellar gas and dust, or with
the Earth's atmosphere, there are
some very intriguing sources like the central parts of Active Galactic
Nuclei (AGN) and some more exotic possibilities like radiation from
nonbaryonic dark matter annihilations and from topological defects. While
a discovery of the latter class of course would be quite remarkable,
also non-discovery is useful to establish limits on the underlying
particle physics theories.

\section{High-Energy Gamma Rays}

Traditionally, gamma ray astronomy has been divided into several 
subfields based on the energy range studied, from the MeV region 
all the way up to YeV ($10^{21}$ eV). This is due to the fact 
that completely different experimental techniques are used, and
also different physical processes are involved in the sources.

In fact, due to the overwhelming background of low-energy gamma rays
produced in the atmosphere by the intense cosmic ray flux, it is necessary
to use space detectors to detect gammas of energy below roughly 50 GeV. 
Above that energy, ground-based air Cherenkov telescopes of much
larger area may be employed. With 
instruments on board the Compton-GRO satellite, notably the EGRET
detector~\cite{egret}, data is now available up to 20 GeV. The EGRET
catalog comprises a large number of supernova remnants and AGNs, but
also many sources of unknown origin. An interesting new result is that the 
diffuse $\gamma$ ray flux
from the galactic center recently  detected by EGRET
seems to show shows some evidence of an excess at high energy which 
is not easily explained in conventional models~\cite{mori}.

At present, there is is an annoying gap in the energy range between
around 20 and 250 GeV, above which energy the most advanced
ground-based air Cherenkov telescopes become functional.  The principle
of these is to detect in optical mirrors the Cherenkov radiation caused by air
showers initiated by the primary particles. Above around 10 TeV, some 
particles of the air showers penetrate all the way down to the surface
(at least at mountain altitudes) and can be detected directly. In air 
shower arrays these cascades are sampled sparsely but over large areas.

The energy gap will most probably be filled from both sides the next 
few years as, e.g., both new space detectors (like GLAST~\cite{glast}) and 
large solar power plant mirror arrays~\cite{ong} are planned to be deployed.

The problem
of establishing a signal from a gamma ray point source is highly nontrivial,
since the cosmic ray flux, roughly $10^{-7}$ cm$^{-2}$ s$^{-1}$ sr$^{-1}$
at 10 TeV, is much higher than any expected gamma ray flux. The low 
signal to noise was probably the reason for some seemingly erroneous
claims of detection of galactic point sources in the 1980's, something
that was rectified by the standard-setting CASA experiment~\cite{casa}. 

The last two or three years, remarkable improvements in the imaging
qualities and hadron rejection of air 
Cherenkov telescopes has finally resulted in solid detection of the 
first few TeV gamma ray point sources. The first one to be detected,
with remarkably high statistics by the Whipple group~\cite{whipcrab},  
was the Crab nebula. 
(It was  confirmed  by several other groups like ASGAT, 
Themistocle, CANGAROO and TIBET.)
The pulsar-driven Crab supernova remnant is 
such a solid TeV gamma ray source that it has become something of a 
standard candle for high-energy gamma ray astronomy today.

The jets 
of AGN had also been hypothesized as being possible TeV gamma ray sources, 
since there is a  Lorentz boost for jets viewed head-on. 
A complication here is that for extragalactic sources, the optical 
depth of gamma rays may become non-negligible. A gamma ray traveling 
through the intergalactic medium will interact with a high cross 
section with photons of energy corresponding to an invariant mass just 
above the $e^+e^-$ cross section. For a TeV photon, this means a sensitivity
to IR photons at $\sim 2\, \mu$m. (Note that the cosmic microwave 
background cuts off PeV $\gamma$ radiation at a fraction of a Mpc.)

 Recently, a detailed analysis~\cite{stecker,macminn} using the most recent 
determinations of the optical and IR intergalactic background,   
has shown that TeV sources more distant than $z\sim 0.1$
 should hardly be seen due to absorption.
 Recent observations seem to  verify this general picture.   

The first observation of TeV gammas from an AGN was made by
the Whipple collaboration~\cite{mrk421}, who detected a signal
from the blazar Mkn 421
at the $6\sigma$ confidence level. Recently, the HEGRA collaboration
independently confirmed this source using two of their 
instruments~\cite{HEGRA421}. Another blazar, Mkn 501, which is too weak a GeV 
source to be detected by EGRET, has recently been seen in TeV 
$\gamma$s by both Whipple and HEGRA~\cite{mrk501}. 

A most remarkable, rapid outburst of TeV $\gamma$s from Mkn 421 was
detected by the Whipple group on May 7th, 1996~\cite{may}. With a doubling 
time of about one hour, the flux increased above the quiescent value
by a factor of more than 50, making this source even brighter
than the Crab in TeV $\gamma$ radiation. In a second outburst about a week 
later, the flux increased by a factor of almost 25 in approximately
30 minutes. This type of violent variability on very short time scales
is bound to severely strain current models, although interesting 
attempts have appeared~\cite{protheroe}.

At this Conference, new results were presented from the HEGRA 
colla\-bora\-tion~\cite{meyer}, indicating that there may be a handful of 
additional TeV sources (in fact, even above 30 TeV) among the nearby
($z \lsim 0.06$) EGRET  sources. If this is 
confirmed, it should have interesting 
consequences for the intergalactic IR and optical background. This 
could give useful information on the mechanisms for early galaxy 
formation~\cite{macminn}.

The origin of the high-energy radiation from  AGNs is still unclear. It 
seems probable that shock acceleration is involved near the black 
hole or, for the blazar class, in the jet, but how particles are 
transferred to the outer regions as well 
as how they interact is still mysterious. In fact, it is not known 
whether leptons or hadrons are mainly responsible for energy 
transport near the accretion region. It is conceivable that 
electrons, interacting with ambient magnetic fields, create 
synchrotron radiation which in turn may be inverse-Compton 
scattered to high energies. These are the so-called SSC (synchrotron
self-Compton) models~\cite{SSC}, which work very successfully
for a supernova remnant like the Crab. In another class of 
models~\cite{Mannheim}, mainly hadrons 
(protons) are accelerated,  which interact with the dense photon gas 
in the AGN central region or in a jet. In $p\gamma\to \pi + X$  
reactions, high-energy neutrinos, electrons, positrons and gamma rays 
are created in the decay of pions. All particles except the neutrinos 
induce electromagnetic cascades which terminate at low energy. In 
particular, the X-ray flux may be used to put an upper bound on the 
neutrino rates in this class of models~\cite{stecker2,venya2}. Although 
estimates are uncertain, it seems that the integrated rate from all 
 AGNs may give a ``diffuse'' source of very high 
energy neutrinos which could be detectable in the new generation of 
neutrino telescopes like AMANDA.

It appears that if the recent detection of $\gamma$s of
more than 30 TeV from several blazars~\cite{meyer} is confirmed, it 
may lend credibility to the 
hadronic model~\cite{beacons}. A solid answer must, however, await a 
detailed analysis of time-correlated multi-waveband data and/or 
the findings from neutrino telescopes.

\section{High-Energy Neutrinos}

Neutrino astronomy was born with the first detection of solar 
neutrinos (too few to 
fit standard solar models) by R. Davis et al.\,in the 1960s,
with the proof two decades later by the Kamiokande collaboration that
the neutrino events really point back to the Sun.
 The solar neutrino problem 
is of course  still one of the  most intriguing  indications
we have for physics outside the Standard Model of particle 
physics~\cite{bahcall}. The remarkable detection of neutrinos from 
SN1987A in the
Kamiokande and IMB detectors (originally constructed to search for 
proton decays) has established neutrino astronomy as a useful branch 
of astrophysics. In addition, the observed neutrino rates from the SN1987A 
event has helped particle physicists to put limits on neutrino 
properties as well as on various hypothetical, weakly interacting 
particles. Indeed, neutrino astrophysics is one of the areas where the 
connections between  astrophysics and particle physics 
are perhaps the strongest. 

The first neutrino telescopes typically had effective areas of the 
order of one to a few hundred m$^2$. They have been followed by a new 
generation (MACRO, Super-Kamiokande) which approaches $10^3$ m$^2$.
Super-Kamiokande, for instance, is an extremely well-equipped and 
sensitive laboratory for all types of neutrino physics of energy from 
a few MeV upwards~\cite{totsuka}. MACRO has recently published~\cite{macro}
 its first measurement of the atmospheric neutrino flux above 1 GeV.

However, for TeV neutrino energies and 
above, all estimates indicate that the effective areas must be much 
larger to give a fair chance of detection~\cite{halzen}. Therefore, a 
new generation of very large telescopes has been developed, 
which sacrifice sensitivity 
of MeV neutrinos for large area ($10^4$ to $10^5$ m$^2$ at present - 
the aim is for 1 km$^2$ within a few years) 
for multi-GeV neutrinos. (Typical thresholds are some tens of
GeV.)  A pioneer of this type was the 
deep ocean DUMAND experiment~\cite{dumand} outside Hawaii, 
which now seems to be discontinued at the prototype 
stage  due to various  technical problems related to the very 
demanding ocean environment. However, even with a small prototype, 
they were able to put some limits on cascades initiated by AGN 
neutrinos~\cite{jeff}, showing the promise of this type
of technique. In Europe, the ocean detector concept is 
being further investigated in the Mediterranean by the 
NESTOR~\cite{nestor} and ANTARES~\cite{antares} collaborations,
 with a large-scale detector still 
being a couple of years ahead.

The Lake Baikal experiment~\cite{baikal} has become the first of the 
natural-water detectors to 
successfully detect atmospheric neutrinos, although only a few 
events so far in its 96-fold OM (optical module) array. The array is 
successively being expanded to 200 OMs, with 3/4 of that expected by
the spring of 1997. It has the 
advantage over ocean detectors of being in fresh water, thus avoiding 
the high radioactive background from $^{40}K$ present in salt water.
Also, the ice cover during winter months helps the logistics of the 
deployment substantially.
However, bioluminescence is present and sedimentation necessitates
regular cleaning of the optical modules. In addition, the relatively shallow 
depth (1300 m) means that a large background of downward atmospheric 
muons has to be fought. In is an impressive achievement of the Baikal
group to have obtained the up/down rejection factor needed to detect
upward-going muons.

In the deep under-ice US-Ger\-man-Swe\-dish de\-tec\-tor AMAN\-DA at 
the South Pole, none of 
these problems is present (although the maximum useable depth
of around 2500 m still gives substantial downward-going muon flux). 
On the other hand, it was not clear before 
last year that the ice quality was good enough to deploy a large 
detector. In particular, a prototype deployed in 1994-95 at 800 to 1000 
m depth showed severe degradation of timing resolution due to 
scattering on residual air bubbles at that depth. However,
ice inbetween air bubbles was found to be remarkably clean, with 
absorption lengths in the near-UV being more than ten times longer 
than ever measured in laboratory ice~\cite{apopt}.

In the 1995-96 season, 4 strings  of 20 OMs  each (20 m 
spacing between OMs)  were deployed 
to 2000 m depth, and the scattering on 
bubbles was found to be absent (or at 
least two orders of magnitude smaller than at 800 m), permitting the 
first muons to be tracked~\cite{hulth}. In the soon finished, highly
successful 1996-97 season, 6 
additional strings have been deployed. Thanks to improvements in 
signal transmission, thinner twisted quad cables could be used 
permitting 36 OMs per string, with now 10 m separation between OMs. 
The average distance between nearest-neighbor strings in the 10-string
detector is around 30 m.
Of the 216 new OMs, only half a dozen have failed,  giving the AMANDA 
collaboration the hope of soon having at its disposal a detector of 
around $10^4$ m$^2$ for upward-going single muons, and much larger for 
cascades initiated, e.g., by electron neutrinos. 

\subsection{Sources of High-Energy Neutrinos}

In a large detector, like the present AMANDA neutrino telescope, 
there will be a real chance to detect neutrinos from AGNs, if the 
models involving acceleration of hadrons are correct. Besides the 
``diffuse'' integrated contribution from all AGNs, which 
could amount to several hundred events per km$^2$ per year~\cite{halzen}, 
the blazars (i.e. AGNs with jets viewed nearly head-on)
from the EGRET catalog will be promising objects to study. The fact 
that the TeV gamma ray sources seen by air Cherenkov telescope are all
relatively nearby, whereas many stronger such EGRET sources are not seen in 
TeV gammas, has as its most natural explanation the intergalactic 
absorption of gamma rays. Thus there could be a large number of very 
intense neutrino sources awaiting discovery.

The fact that whenever hadrons are accelerated, both gamma rays and
neutrinos will be produced through pion decay, means that models,
e.g., for gamma ray bursts (GRBs), where hadronic fireballs are 
excited inevitably predict also neutrino radiation~\cite{halzen2,waxman}. 
In the AMANDA detector, a trigger has been set 
up which can correlate an excess of neutrino events with satellite 
detection of a GRB. (A supernova trigger is also implemented.)
As has been pointed  out~\cite{waxman,learned}, if 
an extragalactic source of neutrinos is found, there are many 
interesting tests of neutrino properties (mass, mixings, magnetic 
moments etc) that can be made, which would supersede terrestrial tests 
and constraints from SN1987A by orders of magnitude.

If  very-high energy (PeV) neutrinos from AGNs are present, a whole
range of other exotic particle physics processes could be investigated 
as well (such as leptoquarks, multi-W processes etc~\cite{blr}). An 
interesting process in addition is the resonant $\bar\nu_{e}+e^-\to W^-$ 
at around 6 PeV, which could give spectacular, background-free 
cascades in Cherenkov detectors.~\cite{gaziz,blr} In fact, for such high 
energies, the way to get a large effective detector volume may be to use 
the coherent radio wave radiation from the shower in the 
ice.~\cite{buford} Prototype radio detectors have been 
deployed piggy-back on AMANDA strings this year.

\subsection{Indirect Detection of Supersymmetric Dark Matter in 
Neutrino Telescopes}

Supersymmetric neutralinos with masses in the GeV--TeV range are among
the leading non-baryonic candidates for the dark matter in our
galactic halo. One of the most promising methods for the discovery of
neutralinos  in
the halo is via observation of energetic neutrinos from their annihilation
in the Sun and/or the Earth~\cite{jkg,bottino,BEG}.  (In some regions
of parameter space, also detection in gamma rays in air Cherenkov 
telescopes through the unique signature
of a line of narrow width, could be feasible~\cite{lbkaplan}.)
 Neutralinos do not
annihilate into neutrinos directly, but energetic
neutrinos may be produced via hadronization and/or
decay  of the direct
annihilation products. These energetic neutrinos may be discovered by
terrestrial neutrino detectors.

The prediction of muon rates is in principle straight-forward but
technically quite involved: one has to compute
neutralino capture rates in the Sun and the Earth, fragmentation
functions in basic annihilation processes, propagation through the
solar or terrestrial medium, charged current cross sections and muon
propagation in the rock, ice or water surrounding the detector. 

The neutralinos $ \tilde{\chi}^0_i$ are linear combinations of the
neutral gauginos ${\tilde B}$, ${\tilde W_3}$ and of the neutral
higgsinos ${\tilde H_1^0}$, ${\tilde H_2^0}$,  
the lightest of which, called $\chi$, is then the candidate for
the particle making up (at least some of) the dark matter in the universe.

With Monte Carlo simulations one can  consider the whole chain of 
processes from the annihilation products in the core of the Sun or the
Earth to detectable muons at the surface of the Earth.

Unfortunately, no details about supersymmetry breaking are known at
present, which means that a lot of parameters are undetermined.
The usual strategy~\cite{bottino,BEG,jkg} is then to scan
the parameter space of the minimal supersymmetric 
extension to the Standard Model. 

The best present limits~\cite{baksan} for indirect searches come from the
Baksan detector. The limits are $\Phi_\mu^{Earth} < 2.1
\times 10^{-14}$ cm$^{-2}$ s$^{-1}$ and $\Phi_\mu^{Sun} < 3.5 \times
10^{-14}$ cm$^{-2}$ s$^{-1}$ at 90\% confidence level and integrated
over a half-angle aperture of 30$^\circ$ with a muon energy threshold
of 1 GeV.  This has already allowed some models to be excluded\cite{BEG}.
A neutrino telescope of an area around 1 km$^2$,
which is a size currently being discussed for a near-future neutrino
telescope, would improve these limits by two or three orders
of magnitude and would have a large
discovery potential for supersymmetric dark matter.

Indirect dark matter searches and LEP2 probe
complementary regions of the supersymmetric parameter space. Moreover,
direct detection~\cite{jkg} is reaching a sensitivity that
allows some models to be excluded\cite{BG}, with somewhat different
characteristics than those probed by the other methods.  This
illustrates a nice complementarity between direct detection, indirect
detection and accelerator methods to bound or confirm the minimal
supersymmetric standard model.

\subsection{Establishing a Neutrino Signal {}from a Point Source}

For neutralino detection, as well as for other physics objectives of 
neutrino telescopes, a problem will always be the irreducible 
background coming from atmospheric neutrinos. However, a typical 
signal will appear as a peak in the angular distribution; usually the 
energy distribution is different as well. The question of how the 
discovery potential depends on the angular and energy resolution has 
recently been investigated~\cite{BEK}.

Due to the finite muon production angle,
one would like to accept muons from a large enough solid angle
around the point source to assure all the
signal events are accepted.  For example, the rms angle between the neutrino
direction and the direction of the induced muon is 
$\sim 20^\circ/\sqrt{E_\nu/10\, {\rm GeV}}$.  
Furthermore, the muon
typically carries half the neutrino energy, so the angular
radius of the acceptance cone should be 
$\sim 14^\circ/\sqrt{E_\mu/10\, {\rm GeV}}$. The problem is of course
that the a priori energy of  signal neutrino events is unknown, so  one
has to optimize angular and energy acceptance according to varying
hypotheses for the neutrino source.

A general covariance-matrix formalism has been  set up~\cite{BEK}  and applied 
to the specific example of neutralino annihilation in the Sun and 
Earth, for detectors with various values of angular and energy resolution.
Comparing, e.g., the improvement by using a 3-parameter fit for
the signal to the simple case of using just one bin up to a certain
angle $\theta_{\rm max}$ one finds
that there could be an improvement of
up to a factor of 2 
at high masses. Although this application was for 
neutralino annihilation, the formalism~\cite{BEK} is general enough to 
be applicable for a generic point source. As large neutrino experiments 
now come on-line, we can expect successive improvements in their 
discovery potential.  
\vskip .1cm
{\bf Conclusions and Acknowledgments}
\vskip .1cm
\noindent
With new windows to the universe, historically it has always been the
case that unexpected discoveries have appeared. I have tried to 
summarize the status and expectations for high-energy gamma ray and 
neutrino astronomy. Maybe the outcome will be different than 
predicted here, but it certainly will be interesting.

The author wishes to thank J.J. Aubert, V. Berezinsky, J. Edsj\"o, 
P.O. Hulth, J. 
Learned, H. Meyer, H. 
Rubinstein, C. Spiering and T. Weekes for useful discussions, and the 
organizers
of ``Texas in Chicago'' for hospitality. This work was sponsored by
the Swedish Natural Science Research Council.

{

\end{document}
\section*{References}}
\begin{thebibliography}{99}
\bibitem{croninetal} J. Cronin, K.G. Gibbs and T.C. Weekes, 
\jou{\annrev}{43}{883}{1993}.
\bibitem{egret} {\texttt http://cossc.gsfc.nasa.gov/cossc/EGRET.html}
\bibitem{mori} M. Mori, Ap. J. {\bf 478} (1997).
\bibitem{glast} GLAST home page: {\texttt http://www-glast.stanford.edu}
\bibitem{ong} R.A. Ong, these Proceedings.
\bibitem{casa} {\texttt http://hep.uchicago.edu/~covault/casa.html}
\bibitem{whipcrab} P.T. Reynolds et al., Astrophys. J. {\bf 404}, 206 (1993).
\bibitem{stecker} F.W. Stecker, O.C. De Jager and M.H. Salamon, Ap. 
J. {\bf 390}, L49 (1992).  
\bibitem{macminn} D. MacMinn and J.R. Primack, Space Sci. Rev. {\bf 75},
413 (1996).
\bibitem{mrk421} M. Punch et al., Nature {\bf 358}, 477 (1992).
\bibitem{HEGRA421} D. Petry et al., Astron. Astrophys. {\bf 311}, L13 (1996).
\bibitem{mrk501} J. Quinn et al., Ap. J. {\bf 456}, L83 (1996); S.M. Bradbury 
et al., astro-ph/9612058 (1996).
\bibitem{may} J.A. Gaidos et al., Nature {\bf 383}, 319 (1996); T.C. Weekes,
these Proceedings.
\bibitem{protheroe} W. Bednarek and R. Protheroe, astro-ph/9612073 
(1996);
 A. Dar and A. Laor, astro-ph/9610252 (1996).
 \bibitem{meyer} H. Meyer, these Proceedings.
 \bibitem{SSC}  C.D. Dermer, R. Schlickeiser and A. Mastichiadis, 
 Astron. Astrophys. {\bf 256}, L27 (1992); A.A. Zdziarski and J.H. 
 Krolik, Ap. J. {\bf 409}, L33 (1994).
 \bibitem{Mannheim} P.L. Biermann and P.A. Strittmatter,
 Ap. J. {\bf 322}, 643 (1987); M.C. Begelman,
 B. Rudak and M. Sikora, Ap. J. {\bf 362}, 38 (1990);  
 K. Mannheim, Phys. Rev. {\bf D48}, 2408 (1993).
\bibitem{stecker2} F.W. Stecker et al., Phys. Rev. Lett. {\bf 66}, 
2697 (1991); (E) {\bf 69}, 2738 (1992).
\bibitem{venya2} V.S. Berezinsky and J.G. Learned, in {\it Proc. of the
Workshop on High Energy Neutrino Astrophysics}, eds. V.J. Stenger, 
J.G. Learned, S. Pakvasa and X. Tata, World Scientific, 1992.
\bibitem{beacons}  K. Mannheim, S. Westerhoff, H. Meyer and H.-H. Fink,
Astron. Astrophys. {\bf 315}, 77 (1996).
\bibitem{bahcall} J.N. Bahcall, these Proceedings.
\bibitem{totsuka} Y. Totsuka, these Proceedings.
\bibitem{macro} S. Ahlen et al., Phys. Lett. {\bf B357} 481.
\bibitem{halzen} T. K. Gaisser, F. Halzen and T. Stanev,
Phys. Rep. {\bf 258}, 173 (1995).
\bibitem{dumand} P.K.F. Grieder, in {\em Trends in Astroparticle Physics}, 
Stockholm,
Sweden, 1994, eds.\ L.~Bergstr\"om, P.~Carlson, P.O.~Hulth and
H.~Snellman, Nucl.\ Phys.\ (Proc.\ Suppl.) {\bf B43} (1995) 265.
\bibitem{jeff} J. Bolesta, these Proceedings.
\bibitem{nestor} L. Resvanis, Europhys. News {\bf 23}, 172 (1992).
\bibitem{antares}ANTARES home page:
{\texttt 
http://marcpl1.in2p3.fr/astro/astro.html}
\bibitem{baikal} I.A. Belolaptikov et al., in {\em Trends in 
Astroparticle Physics}, Stockholm,
Sweden, 1994, eds.\ L.~Bergstr\"om, P.~Carlson, P.O.~Hulth and
H.~Snellman, Nucl.\ Phys.\ (Proc.\ Suppl.) {\bf B43} (1995) 241.
\bibitem{apopt} L. Bergstr\"om et al., physics/9701025, Appl. 
Optics, in press (1997).
\bibitem{hulth} P.O.\ Hulth et al (AMANDA Collaboration),  to appear 
in {\em Proc. of Neutrino 96},  eds. 
K.\ Enqvist, K.\ Huitu and J.\ Maalampi (World Scientific,  Singapore, 1997).
\bibitem{halzen2} F. Halzen and G. Jaczko, astro-ph/9602038 (1996).
\bibitem{waxman} E. Waxman and J.N. Bahcall, astro-ph/9701231 (1997).
\bibitem{learned} J.G. Learned, in {\em Proc. of Neutrino 94}, eds. A. Dar,
G. Eilam and M. Gronau, Nucl. Phys. (Proc. Suppl.) {\bf 38}, 484 (1995).
\bibitem{blr} L. Bergstr\"om, R. Liotta and H. Rubinstein, 
Phys . Lett. {\bf B276}, 231 (1992); N. Arteaga-Romero et al.,
hep-ph/9701339 (1997).
\bibitem{gaziz} V.S. Berezinsky and A.Z. Gazizov, JETP Lett. {\bf 
25}, 254 (1977); R. Gandhi, C. Quigg, M.H. Reno and I. Sarcevic, 
Astropart. Phys. {\bf 5}, 81 (1996).
\bibitem{buford}G. Frichter, J. Ralston and D. Mackay, Phys. Rev. {\bf D53},
1684 (1996); P.B. Price, Astropart. Phys. {\bf 5}, 43 (1996).
\bibitem{jkg} For a comprehensive review containing historical 
references, see G. Jungman, M. Kamionkowski, and K. Griest,
     Phys. Rep. {\bf 267}, 195 (1996).

\bibitem{bottino} V. Berezinsky, A. Bottino, J. Ellis , N. Fornengo,
 G. Mignola and  S. Scopel,  Astropart.Phys. {\bf 5} 333 (1996).

\bibitem{BEG}  L. Bergstr\"om, J. Edsj\"o, and P. Gondolo, hep-ph/9607237,
     Phys.\ Rev.\ {\bf D}, in press (1997).
     
\bibitem{lbkaplan} L. Bergstr\"om and J. Kaplan, Astrop. Phys. {\bf 2},
261 (1994);
G. Jungman and M. Kamionkowski, Phys. Rev. {\bf D51}, 3121 (1995).


\bibitem{baksan} 
M.M. Boliev et al., in {\it TAUP 95}, 
Nucl. Phys. (Proc. Suppl.) 
{\bf B48}, 83  (1996).

\bibitem{BG} L. Bergstr\"om and P. Gondolo, Astropart. Phys. 
{\bf 5}, 263 (1996).

\bibitem{BEK} L. Bergstr\"om, J. Edsj\"o and M. Kamionkowski,
astro-ph/9702037. 



       





\end{thebibliography}
